\documentclass[12pt,english]{article}
\usepackage[T1]{fontenc}
\usepackage[latin9]{inputenc}
\usepackage{amsmath}

\newcommand{\apj}{ApJ}

\newcommand{\mnras}{MNRAS}

\newcommand{\jmp}{J. Math. Phys.}

\usepackage[svgnames]{xcolor}
\usepackage[colorlinks=true,linkcolor=Blue,citecolor=Blue,urlcolor=Blue]{hyperref}

\usepackage{babel}

\title{Perturbation of the Kerr Metric}

\author{Francisco Frutos-Alfaro}
\date{}

\begin{document}

\maketitle

\begin{abstract}
A new Kerr-like metric with quadrupole moment is obtained by means of 
perturbing the Kerr spacetime. By comparison with the exterior Hartle-Thorne 
metric, it is showed that it could be matched to an interior solution. 
This metric may represent the spacetime of an astrophysical object.
\end{abstract}

\section{Introduction \label{sec:00}}

\noindent
In 1963, R.~P.~Kerr \cite{Kerr} proposed a metric that describes a massive 
rotating object. Since then, a huge amount of papers about the structure and 
astrophysical applications of this spacetime appeared. Now, it is widely 
believed that this metric does not represent the spacetime of an astrophysical 
rotating object. This is because the Kerr metric cannot be matched to a 
realistic interior metric \cite{Boshkayev}.  

\noindent
Other multipole and rotating solutions to the Einstein field equations (EFE) 
were obtained by Castej\'on {\it et al.} (1990) \cite{Castejon}, 
Manko \& Novikov (1992) \cite{Manko}, Manko {\it et al.} (2000) \cite{Manko2000}, 
Pachon {\it et al.} (2006) \cite{Pachon}, and Quevedo (1986) 
\cite{Quevedo1986}, Quevedo (1989) \cite{Quevedo1989}, Quevedo \& Mashhoon 
\cite{Quevedo1991}, Quevedo (2011) \cite{Quevedo2011}. In the four first 
articles, they used the Ernst formalism \cite{Ernst}, while in the four last 
ones, the solutions were obtained with the help of 
the Hoenselaers-Kinnersley-Xanthopoulos (HKX) transformations 
\cite{Hoenselaers}. These authors obtain new metrics from a given seed metric. 
These formalisms allow to include other desirable characteristics 
(rotation, multipole moments, magnetic dipole, etc.) to a given seed metrics. 

\noindent
In Nature, it is expected that astrophysical objects are rotating and
slightly deformed. The aim of this article is to derive an appropriate analytical 
tractable metric for calculations in which the quadrupole moment can be 
treated as perturbation, but for arbitrary angular momentum. Moreover, 
this metric should be useful to tackle astrophysical problems, for instance, 
accretion disk in compact stellar objects \cite{Fragile,Hawley}, relativistic 
magnetohydrodynamic jet formation \cite{Fendt}, astrometry \cite{Soffel,Frutos} 
and gravitational lensing \cite{Frutos2001}. Furthermore, software related with 
applications of the Kerr metric can be easily modified in order to include the 
quadrupole moment \cite{Dexter,Vincent,Frutos2012}.
 
\noindent
This paper is organized as follows. In section \ref{sec:01}, we give a succinct 
explanation of the Kerr metric. The weak limit of the Erez-Rosen metric is 
presented in section \ref{sec:02}. In section \ref{sec:03}, the Lewis metric 
is presented. The perturbation method is discussed in section \ref{sec:04}. 
The application of this method leads to a new solution to the EFE with 
quadrupole moment and rotation. It is checked by means of the REDUCE software 
\cite{Hearn} that the resulting metric is solution of the EFE. In section 
\ref{sec:05}, we compare our solution with the exterior Hartle-Thorne metric 
in order to assure that our metric has astrophysical meaning. Forthcoming 
works with this metric are discussed in section \ref{sec:06}.

\section{The Kerr Metric \label{sec:01}}

\noindent
The Kerr metric represents the spacetime of a non-deformed massive rotating 
object. The Kerr metric is given by \cite{Kerr,Carmeli}

\begin{equation}
\label{kerr}
{d} {s}^{2} = \frac{\Delta}{{\rho}^2} [d t - a {\sin}^2 {\theta} d \phi]^2 
- \frac{{\sin}^2 {\theta}}{{\rho}^2} [(r^2 + a^2) d \phi - a d t]^2 
- \frac{{\rho}^2}{\Delta} d {r}^2 - {\rho}^2 d {\theta}^2 , 
\end{equation}

\noindent
where $ \Delta = {r}^2 - 2 M r + {a}^2 $ and 
$ {\rho}^2 = {r}^2 + {a}^2 {\cos}^2 {\theta} $. $ M $ and $ a $ represent the 
mass and the rotation parameter, respectively. The angular momentum of the 
object is $ J = M a $.

\section{The Erez-Rosen metric \label{sec:02}}

\noindent
\noindent 
The Erez-Rosen metric \cite{Carmeli,Winicour,Young,Zeldovich} 
represents a body with quadrupole moment. The principal axis of the quadrupole 
moment is chosen along the spin axis, so that gravitational radiation can be 
ignored. Here, we write down an approximate expression for this metric 
obtained by doing Taylor series \cite{Frutos}

\begin{equation}
\label{erezrosen}
{d} {s}^{2} = \left({1-\frac{2M}{r}}\right) {\rm e}^{-2 \chi} {d} t^{2} 
- \left({1-\frac{2M}{r}}\right)^{-1} {\rm e}^{2 \chi} {{d} r^{2}} 
- {r^{2}}{\rm e}^{2 \chi} ({d} {\theta}^{2} + \sin^2{\theta} {d} {\phi}^{2}) ,
\end{equation}

\noindent
where $ {d} {\Sigma}^2 =  {d} {\theta}^{2} + \sin^2{\theta} {d} {\phi}^{2} $, 
and 

\begin{equation}
\label{chi}
\chi = \frac{2}{15} q \frac{M^{3}}{r^{3}} P_{2}(\cos{\theta}) .
\end{equation}

\noindent
The quadrupole parameter is given by $ q = 15 G Q / (2 c^{2} M^{3}) $, with 
$ Q $ representing the quadrupole moment. This metric is valid up to the order 
$ O(q M^4, \, q^2) $.

\section{The Lewis Metrics \label{sec:03}}

\noindent 
The Lewis metric is given by \cite{Lewis,Carmeli} 

\begin{equation}
\label{lewis} 
{d}{s}^2 = V d t^2 - 2 W d t d \phi 
- {\rm e}^{\mu} d \rho^2 - {\rm e}^{\nu} d z^2 - Z d \phi^2
\end{equation}

\noindent 
where we have chosen the canonical co\-or\-di\-na\-tes $ x^{1} = \rho $ and 
$ x^{2} = z $, $ V, \, W, \, Z $, $ \mu $ and $ \nu $ are functions of $ \rho $ 
and $ z $ ($ \rho^2 = V Z + W^2 $). Choosing $ \mu = \nu $ and performing the 
following changes of potentials

$$ V = f , \quad W = \omega f , \quad Z = \frac{\rho^2}{f} - \omega^2 f 
\quad {\rm and} \quad {\rm e}^{\mu} = \frac{{\rm e}^{\gamma}}{f} , $$

\noindent 
we get the Papapetrou metric

\begin{equation}
\label{papapetrou} 
{d}{s}^2 = f (d t - \omega d \phi)^2 
- \frac{{\rm e}^{\gamma}}{f} [d \rho^2 + d z^2] - \frac{\rho^2}{f} d \phi^2 .
\end{equation}

\section{Perturbing the Kerr Metric \label{sec:04}}

\noindent
To include a small quadrupole moment into the Kerr metric we will modify the 
Lewis-Pa\-pa\-pe\-trou metric (\ref{papapetrou}). First of all, we choose 
expressions for the canonical coordinates $ \rho $ and $ z $. For the Kerr 
metric \cite{Kerr}, one particular choice is \cite{Carmeli,Chandrasekhar} 

\begin{equation}
\label{chandra} 
\rho = \sqrt{\Delta} \sin{\theta} \qquad {\rm and} \qquad 
z = (r - M) \cos{\theta} 
\end{equation}

\noindent 
where $ \Delta = r^2 - 2 M r + a^2 $.

\noindent 
From (\ref{chandra}) we get

\begin{equation}
\label{cylindric}
d \rho^2 + d z^2 = 
[ {(r - M)^2} \sin^2{\theta} + \Delta \cos^2{\theta} ] 
\left( \frac{d r^2}{\Delta} + d {\theta}^2 \right) .
\end{equation}

\noindent 
If we choose

$$ {{\rm e}^{\mu}} 
= {\tilde{\rho}}^2 [ {(r - M)^2} \sin^2{\theta} + \Delta \cos^2{\theta} ]^{-1} , $$
 
\noindent 
the term (\ref{cylindric}) becomes

$$ e^{\mu} [ d \rho^2 + d z^2 ] = 
{\tilde{\rho}}^2 \left(\frac{{{d} r^{2}}}{\Delta} + {d} {\theta}^{2} \right) , $$

\noindent
where $ {\tilde{\rho}}^2 = r^2 + a^2 \cos^2{\theta} $.

\noindent 
From (\ref{papapetrou}), we propose the following metric 

\begin{equation}
\label{lewis2} 
{d}{s}^2 = {\cal V} d t^2 - 2 {\cal W} d t d \phi - {\cal X} {{d} r^{2}} 
- {\cal Y} {d} \theta^{2} - {\cal Z} {d} \phi^{2} , 
\end{equation}

\noindent 
where

\begin{eqnarray}
{\cal V} & = & V {\rm e}^{- 2 \psi} \nonumber \\
{\cal W} & = & W \nonumber \\
{\cal X} & = & X {\rm e}^{2 \psi} \\
{\cal Y} & = & Y {\rm e}^{2 \psi} \nonumber \\
{\cal Z} & = & Z {\rm e}^{2 \psi} , \nonumber  
\end{eqnarray}

\noindent 
where the potentials $ V, \, W, \, X, \, Y, \, Z $, and $ \psi $ depend on 
$ x^1 = r $ and $ x^2 = \theta $.

\noindent 
Now, let us choose 

\begin{eqnarray}
V & = & f = \frac{1}{{\tilde{\rho}}^2} [\Delta - a^2 \sin^2{\theta}] \nonumber \\ 
W & = & \frac{a}{{\tilde{\rho}}^2} [\Delta - (r^2 + a^2)] \sin^2{\theta} 
= - \frac{2 J r}{{\tilde{\rho}}^2} \sin^2{\theta} \nonumber \\
X & = & \frac{{\tilde{\rho}}^2}{\Delta} \\
Y & = & {\tilde{\rho}}^2  \nonumber \\
Z & = & \frac{\sin^2{\theta}}{{\tilde{\rho}}^2} 
[(r^2 + a^2)^2 - a^2 \Delta \sin^2{\theta}] . \nonumber 
\end{eqnarray}

\noindent
The only potential we have to find is $ \psi $. In order to obtain this 
potential, the EFE must be solved 

\begin{equation}
\label{einstein} 
G_{i j} = R_{i j} - \frac{R}{2} g_{i j} = 0 
\end{equation}

\noindent 
where $ R_{i j} $ ($ i, \, j = 0, \, 1,\, 2, \, 3 $) are the Ricci tensor 
components and $ R $ is the curvature scalar. The Ricci tensor components and 
the curvature scalar $ R $ for this metric can be found in the Appendix. 

\noindent 
In our calculations, we consider the potential $ \psi $ as perturbation, 
{\it i.e.} one neglects terms of the form 

$$ \left(\frac{\partial \psi}{\partial r} \right)^2 = 
\left(\frac{\partial \psi}{\partial \theta} \right)^2 = 
\frac{\partial \psi}{\partial r} \frac{\partial \psi}{\partial \theta}  
\sim 0 . $$ 

% $ q^2 = a q = m q \simeq 0 $ 

\noindent 
Terms containing factors of the form 

$$ a \frac{\partial \psi}{\partial x^i} 
= m \frac{\partial \psi}{\partial x^i} \sim 0 
\qquad (i = 1, \, 2) $$

\noindent 
are also neglected. Substituting the known potentials 
($ V, \, W, \, X, \, Y, \, Z $) into the expressions for 
the Ricci tensor and the curvature scalar (see Appendix), it results only one 
equation for $ \psi $ that we have to solved:

\begin{equation}
\label{eqdif}
\sin{\theta} \frac{\partial}{\partial r} 
\left(r^2 \frac{\partial \psi}{\partial r} \right) 
+ \frac{\partial}{\partial \theta} \left(\sin{\theta} 
\frac{\partial \psi}{\partial \theta} \right) = 0 
\end{equation}

\noindent 
The solution for this equation is

\begin{equation}
\label{solution}
\psi = \frac{\cal K}{r^3} P_{2} (\cos{\theta}) , 
\end{equation}

\noindent 
where $ {\cal K} $ is a constant. To determine this constant, we compare the 
weak limit of the metric (\ref{lewis2}) with the Erez-Rosen metric 
(\ref{erezrosen}), {\it i.e.} $ \psi = \chi $. The result is 
$ {\cal K} = 2 q M^3 / 15 $.
  
\noindent 
Then, the new modified Kerr metric containing quadrupole moment is

\begin{eqnarray}
\label{newkerr}
{d} {s}^{2} & = & \frac{{\rm e}^{- 2 \chi}}{{\rho}^2}
[\Delta - {a}^2 {\sin}^2 {\theta}] d {t}^2 
+ \frac{4 J r}{{\rho}^2} {\sin}^2 {\theta} d {t} d {\phi} 
- \frac{{\rho}^2 {\rm e}^{2 \chi}}{\Delta} d {r}^2 
- {\rho}^2 {\rm e}^{2 \chi} d {\theta}^2 \nonumber \\
& - & \frac{{\rm e}^{2 \chi} {\sin}^2 {\theta}}{{\rho}^2} 
[({r}^2 + {a}^2)^2 - {a}^2 \Delta {\sin}^2 {\theta}] d {\phi}^2 \nonumber \\
& = & \frac{\Delta}{{\rho}^2} [{\rm e}^{- \chi} d t 
- a {\rm e}^{\chi} {\sin}^2 {\theta} d \phi]^2 
- \frac{{\sin}^2 {\theta}}{{\rho}^2} [(r^2 + a^2) {\rm e}^{\chi} d \phi 
- a {\rm e}^{- \chi} d t]^2 
\nonumber \\
& - & {\rm e}^{2 \chi} \left(\frac{{\rho}^2}{\Delta} d {r}^2 
+ {\rho}^2 d {\theta}^2 \right) , 
\end{eqnarray}

\noindent 
where the tilde over the $ \rho $ is dropped.

\noindent 
We verified that the metric \eqref{newkerr} is indeed a solution of the EFE 
using REDUCE \cite{Hearn} up to the order $ O(q M^4, \, q^2) $.

\section{Comparison with the Exterior Hartle-Thorne Metric \label{sec:05}}

\noindent 
In order to establish whether the metric (\ref{newkerr}) does really represent 
the gravitational field of an astrophysical object, we should show that it is 
possible to construct an interior solution, which can appropriately be matched 
with the exterior solution. For this purpose, Boshkayev {\it et al.} 
\cite{Boshkayev} and Frutos-Alfaro {\it et al.} \cite{Frutos} employed 
the exterior Hartle-Thorne metric \cite{Hartle,Berti}

\begin{eqnarray}
\label{hartle}
d s^2 & = & 
\left(1 - \frac{2 {\cal M}}{r} 
+ \frac{2 {\cal Q} {\cal M}^3}{r^3} P_2(\cos{\theta}) \right) d t^2 \nonumber \\
& - & \left(1 + \frac{2 {\cal M}}{r} + \frac{4 {\cal M}^2}{r^2} 
- \frac{2 {\cal Q} {\cal M}^3 }{r^3} P_2(\cos{\theta}) \right) d r^2 \\
& - & r^2 \left(1 - \frac{2 {\cal Q} {\cal M}^3}{r^3} P_2(\cos{\theta}) \right) 
d \Sigma^2 + \frac{4 {\cal J}}{r} \sin^2{\theta} d t d \phi , 
\nonumber  
\end{eqnarray}

\noindent 
where $ {\cal M} $, $ {\cal J} $, and $ {\cal Q} $ are related with the total 
mass, angular momentum, and mass quadrupole moment of the rotating object, 
respectively.

\noindent 
The spacetime (\ref{newkerr}) has the same weak limit as the metric obtained 
by Frutos {\it et al.} \cite{Frutos}. A comparison of the exterior 
Hartle-Thorne metric \cite{Hartle} with the weak limit of the metric 
(\ref{newkerr}) shows that upon defining 

\begin{equation}
\label{definitions}
{\cal M} = M, \qquad {\cal J} = J, \qquad 
2 {\cal Q} {\cal M}^3 = - \frac{4}{15} q {M^3} , 
\end{equation}

\noindent 
both metrics coincide up to the order $ O(M^3, \, a^2, \, q M^4, \, q^2) $. 
Hence, the metric (\ref{newkerr}) may be used to represent a compact 
astrophysical object.

%\section{Applications of the Metric \label{sec:06}}

%\noindent

\section{Conclusions \label{sec:06}}

\noindent
The new Kerr metric with quadrupole moment was obtained by solving the 
EFE approximately. It may represent the spacetime of a rotating and slightly 
deformed astrophysical object. This is possible, because it could be matched 
to an interior solution. We showed it by comparison of our metric with the 
exterior Hartle-Thorne metric. Moreover, the inclusion of the quadrupole moment 
in the Kerr metric does it more suitable for astrophysical calculations than 
the Kerr metric alone. There are a large variety of applications which can be 
tackled with this new metric. Amongst the applications for this metric are 
astrometry, gravitational lensing, relativistic magnetohydrodynamic jet formation, 
and accretion disks in compact stellar objects. Furthermore, the existing software 
with applications of the Kerr metric can be easily modified to include 
the quadrupole moment.

\appendix
\section{Appendix}

\begin{eqnarray*}
R_{0 0} & = & \frac{{\rm e}^{- 2 \psi}}{4 \rho^2 X^2 Y^2} \left( 
- 4 \rho^2 V X^2 Y \frac{\partial^2 \psi}{\partial \theta^2} 
+ 8 V W^2 X^2 Y \left(\frac{\partial \psi}{\partial \theta} \right)^2 \right. \\
& - & \left. 2 \rho^2 V X Y 
\frac{\partial \psi}{\partial \theta} \frac{\partial X}{\partial \theta} 
+ 2 V X^2 Y \frac{\partial \psi}{\partial \theta} 
\frac{\partial \rho^2}{\partial \theta} 
- 4 \rho^2 X^2 Y \frac{\partial \psi}{\partial \theta} 
\frac{\partial V}{\partial \theta} \right. \\
& - & \left. 4 W^2 X^2 Y \frac{\partial \psi}{\partial \theta} 
\frac{\partial V}{\partial \theta} 
+ 2 \rho^2 V X^2  \frac{\partial \psi}{\partial \theta} 
\frac{\partial Y}{\partial \theta} 
- 4 V^2 X^2 Y \frac{\partial \psi}{\partial \theta} 
\frac{\partial Z}{\partial \theta} \right. \\
& - & \left. 4 \rho^2 V X Y^2 \frac{\partial^2 \psi}{\partial r^2} 
+ 8 V W^2 X Y^2 \left(\frac{\partial \psi}{\partial r} \right)^2 
+ 2 \rho^2 V Y^2 \frac{\partial \psi}{\partial r} \frac{\partial X}{\partial r} 
\right. \\
& + & \left. 
2 V X Y^2 \frac{\partial \psi}{\partial r} \frac{\partial \rho^2}{\partial r} 
- 4 \rho^2 X Y^2 \frac{\partial \psi}{\partial r} \frac{\partial V}{\partial r} 
- 4 W^2 X Y^2 \frac{\partial \psi}{\partial r} \frac{\partial V}{\partial r} 
\right. \\
& - & \left.
2 \rho^2 V X Y \frac{\partial \psi}{\partial r} \frac{\partial Y}{\partial r} 
- 4 V^2 X Y^2 \frac{\partial \psi}{\partial r} \frac{\partial Z}{\partial r} 
+ \rho^2 X Y \frac{\partial X}{\partial \theta} 
\frac{\partial V}{\partial \theta} \right. \\
& - & \left. 
\rho^2 Y^2 \frac{\partial X}{\partial r} \frac{\partial V}{\partial r} 
- X^2 Y \frac{\partial \rho^2}{\partial \theta} 
\frac{\partial V}{\partial \theta} 
- X Y^2 \frac{\partial \rho^2}{\partial r} \frac{\partial V}{\partial r} 
\right. \\
& + & \left. 2 \rho^2 X^2 Y \frac{\partial^2 V}{\partial \theta^2} 
- \rho^2 X^2 \frac{\partial V}{\partial \theta} 
\frac{\partial Y}{\partial \theta} 
+ 2 V X^2 Y \frac{\partial V}{\partial \theta} 
\frac{\partial Z}{\partial \theta} \right. \\
& + & \left. 2 \rho^2 X Y^2 \frac{\partial^2 V}{\partial r^2} 
+ \rho^2 X Y \frac{\partial V}{\partial r} \frac{\partial Y}{\partial r} 
+ 2 V X Y^2 \frac{\partial V}{\partial r} \frac{\partial Z}{\partial r} 
\right. \\
& + & \left. 2 V X^2 Y \left(\frac{\partial W}{\partial \theta} \right)^2 
+ 2 V X Y^2 \left(\frac{\partial W}{\partial r} \right)^2 \right) \\
R_{0 1} & = & 0 \\
R_{0 2} & = & 0
\end{eqnarray*}

\newpage

\begin{eqnarray*}
R_{0 3} & = & \frac{{\rm e}^{- 2 \psi}}{4 \rho^2 X^2 Y^2} \left(
8 \rho^2 W X^2 Y \left(\frac{\partial \psi}{\partial \theta} \right)^2 
- 8 W^3 X^2 Y \left(\frac{\partial \psi}{\partial \theta} \right)^2 \right. \\
& - & \left. 4 W X^2 Y \frac{\partial \psi}{\partial \theta} 
\frac{\partial \rho^2}{\partial \theta} 
+ 8 W^2 X^2 Y \frac{\partial \psi}{\partial \theta} 
\frac{\partial W}{\partial \theta} 
+ 8 V W X^2 Y \frac{\partial \psi}{\partial \theta} 
\frac{\partial Z}{\partial \theta} \right. \\
& + & \left. 8 \rho^2 W X Y^2 \left(\frac{\partial \psi}{\partial r} \right)^2 
- 8 W^3 X Y^2 \left(\frac{\partial \psi}{\partial r} \right)^2 
- 4 W X Y^2 \frac{\partial \psi}{\partial r} \frac{\partial \rho^2}{\partial r} 
\right. \\
& + & \left. 
8 W^2 X Y^2 \frac{\partial \psi}{\partial r} \frac{\partial W}{\partial r} 
+ 8 V W X Y^2 \frac{\partial \psi}{\partial r} \frac{\partial Z}{\partial r} 
- \rho^2 X Y \frac{\partial X}{\partial \theta} 
\frac{\partial W}{\partial \theta} \right. \\
& + & \left. 
\rho^2 Y^2 \frac{\partial X}{\partial r} \frac{\partial W}{\partial r} 
+ X^2 Y \frac{\partial \rho^2}{\partial \theta} 
\frac{\partial W}{\partial \theta} 
+ X Y^2 \frac{\partial \rho^2}{\partial r} \frac{\partial W}{\partial r} 
\right. \\
& - & \left. 2 W X^2 Y \frac{\partial V}{\partial \theta} 
\frac{\partial Z}{\partial \theta} 
- 2 W X Y^2 \frac{\partial V}{\partial r} \frac{\partial Z}{\partial r} 
- 2 \rho^2 X^2 Y \frac{\partial^2 W}{\partial \theta^2} \right. \\
& - & \left. 2 W X^2 Y \left(\frac{\partial W}{\partial \theta} \right)^2 
+ \rho^2 X^2 \frac{\partial W}{\partial \theta} 
\frac{\partial Y}{\partial \theta} 
- 2 \rho^2 X Y^2 \frac{\partial^2 W}{\partial r^2} \right. \\ 
& - & \left. 2 W X Y^2 \left(\frac{\partial W}{\partial r} \right)^2 
- \rho^2 X Y \frac{\partial W}{\partial r} \frac{\partial Y}{\partial r} 
\right)
\end{eqnarray*}

\newpage 

\begin{eqnarray*}
R_{1 1} & = & \frac{1}{4 \rho^4 X Y^2} \left( 
- 4 \rho^4 X^2 Y \frac{\partial^2 \psi}{\partial \theta^2} 
- 2 \rho^4 X Y \frac{\partial \psi}{\partial \theta} 
\frac{\partial X}{\partial \theta} \right. \\
& - & \left. 2 \rho^2 X^2 Y \frac{\partial \psi}{\partial \theta} 
\frac{\partial \rho^2}{\partial \theta} 
+ 2 \rho^4 X^2 \frac{\partial \psi}{\partial \theta} 
\frac{\partial Y}{\partial \theta} 
- 4 \rho^4 X Y^2 \frac{\partial^2 \psi}{\partial r^2} \right. \\
& - & \left. 8 \rho^4 X Y^2 \frac{\partial \psi}{\partial r}^2 
+ 8 \rho^2 W^2 X Y^2 \left(\frac{\partial \psi}{\partial r} \right)^2 
+ 2 \rho^4 Y^2 \frac{\partial \psi}{\partial r} \frac{\partial X}{\partial r} 
\right. \\
& + & \left. 6 \rho^2 X Y^2 \frac{\partial \psi}{\partial r} 
\frac{\partial \rho^2}{\partial r} 
- 8 \rho^2 W X Y^2 \frac{\partial \psi}{\partial r} 
\frac{\partial W}{\partial r} 
- 2 \rho^4 X Y \frac{\partial \psi}{\partial r} \frac{\partial Y}{\partial r} 
\right. \\
& - & \left. 8 \rho^2 V X Y^2 \frac{\partial \psi}{\partial r} 
\frac{\partial Z}{\partial r} 
- 2 \rho^4 X Y \frac{\partial^2 X}{\partial \theta^2} 
+ \rho^4 Y \left(\frac{\partial X}{\partial \theta} \right)^2 \right. \\
& - & \left. \rho^2 X Y \frac{\partial X}{\partial \theta} 
\frac{\partial \rho^2}{\partial \theta} 
+ \rho^4 X \frac{\partial X}{\partial \theta} 
\frac{\partial Y}{\partial \theta} 
+ \rho^2 Y^2 \frac{\partial X}{\partial r} \frac{\partial \rho^2}{\partial r} 
\right. \\
& + & \left. 
\rho^4 Y \frac{\partial X}{\partial r} \frac{\partial Y}{\partial r} 
- 2 \rho^2 X Y^2 \frac{\partial^2 \rho^2}{\partial r^2} 
+ X Y^2 \left(\frac{\partial \rho^2}{\partial r} \right)^2 \right. \\
& + & \left. 
2 V X Y^2 \frac{\partial \rho^2}{\partial r} \frac{\partial Z}{\partial r} 
+ 2 W^2 X Y^2 \frac{\partial V}{\partial r} \frac{\partial Z}{\partial r} 
+ 2 \rho^2 X Y^2 \left(\frac{\partial W}{\partial r} \right)^2 \right. \\
& - & \left.
 4 V W X Y^2 \frac{\partial W}{\partial r} \frac{\partial Z}{\partial r} 
- 2 \rho^4 X Y \frac{\partial^2 Y}{\partial r^2} 
+ \rho^4 X \left(\frac{\partial Y}{\partial r} \right)^2 \right. \\
& - & \left. 2 V^2 X Y^2 \left(\frac{\partial Z}{\partial r} \right)^2 \right)
\end{eqnarray*}

\newpage

\begin{eqnarray*}
R_{1 2} & = & \frac{1}{4 \rho^4 X Y} \left( 
- 8 \rho^4 X Y \frac{\partial \psi}{\partial \theta} 
\frac{\partial \psi}{\partial r} 
+ 8 \rho^2 W^2 X Y \frac{\partial \psi}{\partial \theta} 
\frac{\partial \psi}{\partial r} \right. \\
& + & \left. 
4 \rho^2 X Y \frac{\partial \psi}{\partial \theta} 
\frac{\partial \rho^2}{\partial r} 
- 4 \rho^2 W X Y \frac{\partial \psi}{\partial \theta} 
\frac{\partial W}{\partial r} 
- 4 \rho^2 V X Y \frac{\partial \psi}{\partial \theta} 
\frac{\partial Z}{\partial r} \right. \\
& + & \left. 4 \rho^2 X Y \frac{\partial \psi}{\partial r} 
\frac{\partial \rho^2}{\partial \theta} 
- 4 \rho^2 W X Y \frac{\partial \psi}{\partial r} 
\frac{\partial W}{\partial \theta} 
- 4 \rho^2 V X Y \frac{\partial \psi}{\partial r} 
\frac{\partial Z}{\partial \theta} \right. \\
& + & \left. \rho^2 Y \frac{\partial X}{\partial \theta} 
\frac{\partial \rho^2}{\partial r} 
- 2 \rho^2 X Y \frac{\partial^2 \rho^2}{\partial \theta \partial r} 
+ W^2 X Y \frac{\partial^2 \rho^2}{\partial \theta \partial r} \right. \\
& + & \left. X Y \frac{\partial \rho^2}{\partial \theta} 
\frac{\partial \rho^2}{\partial r} 
+ \rho^2 X \frac{\partial \rho^2}{\partial \theta} 
\frac{\partial Y}{\partial r} 
+ V X Y \frac{\partial \rho^2}{\partial \theta} \frac{\partial Z}{\partial r} 
\right. \\
& + & \left. 
V X Y \frac{\partial \rho^2}{\partial r} \frac{\partial Z}{\partial \theta} 
- W^2 X Y Z \frac{\partial^2 V}{\partial \theta \partial r} 
- 2 W^3 X Y \frac{\partial^2 W}{\partial \theta \partial r} \right. \\
& + & \left. 
2 \rho^2 X Y \frac{\partial W}{\partial \theta} \frac{\partial W}{\partial r} 
- 2 W^2 X Y \frac{\partial W}{\partial \theta} \frac{\partial W}{\partial r} 
- 2 V W X Y \frac{\partial W}{\partial \theta} \frac{\partial Z}{\partial r} 
\right. \\
& - & \left. 
2 V W X Y \frac{\partial W}{\partial r} \frac{\partial Z}{\partial \theta} 
- V W^2 X Y \frac{\partial^2 Z}{\partial \theta \partial r} 
- 2 V^2 X Y \frac{\partial Z}{\partial \theta} \frac{\partial Z}{\partial r} 
\right) \\
R_{1 3} & = & 0
\end{eqnarray*}

\newpage

\begin{eqnarray*}
R_{2 2} & = & \frac{1}{4 \rho^4 X^2 Y} \left( 
- 4 \rho^4 X^2 Y \frac{\partial^2 \psi}{\partial \theta^2} 
- 8 \rho^4 X^2 Y \left(\frac{\partial \psi}{\partial \theta} \right)^2 
\right. \\
& + & \left. 
8 \rho^2 W^2 X^2 Y \left(\frac{\partial \psi}{\partial \theta} \right)^2 
- 2 \rho^4 X Y \frac{\partial \psi}{\partial \theta} 
\frac{\partial X}{\partial \theta} 
+ 6 \rho^2 X^2 Y \frac{\partial \psi}{\partial \theta} 
\frac{\partial \rho^2}{\partial \theta} \right. \\
& - & \left. 8 \rho^2 W X^2 Y \frac{\partial \psi}{\partial \theta} 
\frac{\partial W}{\partial \theta} 
+ 2 \rho^4 X^2 \frac{\partial \psi}{\partial \theta} 
\frac{\partial Y}{\partial \theta} 
- 8 \rho^2 V X^2 Y \frac{\partial \psi}{\partial \theta} 
\frac{\partial Z}{\partial \theta} \right. \\
& - & \left. 4 \rho^4 X Y^2 \frac{\partial^2 \psi}{\partial r^2} 
+ 2 \rho^4 Y^2 \frac{\partial \psi}{\partial r} \frac{\partial X}{\partial r} 
- 2 \rho^2 X Y^2 \frac{\partial \psi}{\partial r} 
\frac{\partial \rho^2}{\partial r} \right. \\
& - & \left. 
2 \rho^4 X Y \frac{\partial \psi}{\partial r} \frac{\partial Y}{\partial r} 
- 2 \rho^4 X Y \frac{\partial^2 X}{\partial \theta^2} 
+ \rho^4 Y \left(\frac{\partial X}{\partial \theta} \right)^2 \right. \\
& + & \left. \rho^4 X \frac{\partial X}{\partial \theta} 
\frac{\partial Y}{\partial \theta} 
+ \rho^4 Y \frac{\partial X}{\partial r} \frac{\partial Y}{\partial r} 
- 2 \rho^2 X^2 Y \frac{\partial^2 \rho^2}{\partial \theta^2} \right. \\
& + & \left. X^2 Y \left(\frac{\partial \rho^2}{\partial \theta} \right)^2 
+ \rho^2 X^2 \frac{\partial \rho^2}{\partial \theta} 
\frac{\partial Y}{\partial \theta} 
+ 2 V X^2 Y \frac{\partial \rho^2}{\partial \theta} 
\frac{\partial Z}{\partial \theta} \right. \\
& - & \left. 
\rho^2 X Y \frac{\partial \rho^2}{\partial r} \frac{\partial Y}{\partial r} 
+ 2 W^2 X^2 Y \frac{\partial V}{\partial \theta} 
\frac{\partial Z}{\partial \theta} 
+ 2 \rho^2 X^2 Y \left(\frac{\partial W}{\partial \theta} \right)^2 \right. \\
& - & \left. 4 V W X^2 Y \frac{\partial W}{\partial \theta} 
\frac{\partial Z}{\partial \theta} 
- 2 \rho^4 X Y \frac{\partial^2 Y}{\partial r^2} 
+ \rho^4 X \left(\frac{\partial Y}{\partial r} \right)^2 \right. \\ 
& - & \left. 
2 V^2 X^2 Y \left(\frac{\partial Z}{\partial \theta} \right)^2 \right) \\ 
R_{2 3} & = & 0
\end{eqnarray*}

\newpage

\begin{eqnarray*}
R_{3 3} & = & \frac{1}{4 \rho^2 X^2 Y^2} \left( 
- 4 \rho^2 X^2 Y Z \frac{\partial^2 \psi}{\partial \theta^2} 
- 8 W^2 X^2 Y Z \left(\frac{\partial \psi}{\partial \theta} \right)^2 
\right. \\  
& - & \left. 2 \rho^2 X Y Z \frac{\partial \psi}{\partial \theta} 
\frac{\partial X}{\partial \theta} 
- 2 Y X^2 Z \frac{\partial \psi}{\partial \theta} 
\frac{\partial \rho^2}{\partial \theta} 
+ 8 W X^2 Y Z \frac{\partial \psi}{\partial \theta} 
\frac{\partial W}{\partial \theta} \right. \\ 
& + & \left. 2 \rho^2 X^2 Z \frac{\partial \psi}{\partial \theta} 
\frac{\partial Y}{\partial \theta} 
- 8 W^2 X^2 Y \frac{\partial \psi}{\partial \theta} 
\frac{\partial Z}{\partial \theta} 
- 4 \rho^2 X Y^2 Z \frac{\partial^2 \psi}{\partial r^2} \right. \\ 
& - & \left. 8 W^2 X Y^2 Z \left(\frac{\partial \psi}{\partial r} \right)^2 
+ 2 \rho^2 Y^2 Z \frac{\partial \psi}{\partial r} \frac{\partial X}{\partial r} 
- 2 X Y^2 Z \frac{\partial \psi}{\partial r} \frac{\partial \rho^2}{\partial r} 
\right. \\ 
& + & \left. 
8 W X Y^2 Z \frac{\partial \psi}{\partial r} \frac{\partial W}{\partial r} 
- 2 \rho^2 X Y Z \frac{\partial \psi}{\partial r} \frac{\partial Y}{\partial r} 
- 8 W^2 X Y^2 \frac{\partial \psi}{\partial r} \frac{\partial Z}{\partial r} 
\right. \\ 
& - & \left. \rho^2 X Y \frac{\partial X}{\partial \theta} 
\frac{\partial Z}{\partial \theta} 
+ \rho^2 Y^2 \frac{\partial X}{\partial r} \frac{\partial Z}{\partial r} 
- X^2 Y \frac{\partial \rho^2}{\partial \theta} 
\frac{\partial Z}{\partial \theta} \right. \\ 
& - & \left. 
X Y^2 \frac{\partial \rho^2}{\partial r} \frac{\partial Z}{\partial r} 
- 2 X^2 Y Z \left(\frac{\partial W}{\partial \theta} \right)^2 
+ 4 W X^2 Y \frac{\partial W}{\partial \theta} 
\frac{\partial Z}{\partial \theta} \right. \\ 
& - & \left. 2 X Y^2 Z \left(\frac{\partial W}{\partial r} \right)^2 
+ 4 W X Y^2 \frac{\partial W}{\partial r} \frac{\partial Z}{\partial r} 
+ \rho^2 X^2 \frac{\partial Y}{\partial \theta} 
\frac{\partial Z}{\partial \theta} \right. \\ 
& - & \left. 
\rho^2 X Y \frac{\partial Y}{\partial r} \frac{\partial Z}{\partial r} 
- 2 \rho^2 X^2 Y \frac{\partial^2 Z}{\partial \theta^2} 
+ 2 V X^2 Y \left(\frac{\partial Z}{\partial \theta} \right)^2 \right. \\ 
& - & \left. 2 \rho^2 X Y^2 \frac{\partial^2 Z}{\partial r^2} 
+ 2 V X Y^2 \left(\frac{\partial Z}{\partial r} \right)^2 \right)
\end{eqnarray*}

\newpage

\noindent
Calculation of the scalar curvature

\begin{eqnarray*}
R & = & \frac{{\rm e}^{- 2 \psi} }{2 \rho^4 X^2 Y^2} \left(
4 \rho^4 X^2 Y \frac{\partial^2 \psi}{\partial \theta^2} 
+ 4 \rho^4 X^2 Y \left(\frac{\partial \psi}{\partial \theta} \right)^2 
\right. \\ 
& - & \left. 
4 \rho^2 W^2 X^2 Y \left(\frac{\partial \psi}{\partial \theta} \right)^2 
+ 2 \rho^4 X Y \frac{\partial \psi}{\partial \theta} 
\frac{\partial X}{\partial \theta} 
- 2 \rho^2 X^2 Y \frac{\partial \psi}{\partial \theta} 
\frac{\partial \rho^2}{\partial \theta} \right. \\ 
& + & \left. 4 \rho^2 W X^2 Y \frac{\partial \psi}{\partial \theta} 
\frac{\partial W}{\partial \theta} 
- 2 \rho^4 X^2 \frac{\partial \psi}{\partial \theta} 
\frac{\partial Y}{\partial \theta} 
+ 4 \rho^2 V X^2 Y \frac{\partial \psi}{\partial \theta} 
\frac{\partial Z}{\partial \theta} \right. \\ 
& + & \left. 4 \rho^4 X Y^2 \frac{\partial^2 \psi}{\partial r^2} 
+ 4 \rho^4 X Y^2 \left(\frac{\partial \psi}{\partial r} \right)^2 
- 4 \rho^2 W^2 X Y^2 \left(\frac{\partial \psi}{\partial r} \right)^2 
\right. \\ 
& - & \left. 
2 \rho^4 Y^2 \frac{\partial \psi}{\partial r} \frac{\partial X}{\partial r} 
- 2 \rho^2 X Y^2 \frac{\partial \psi}{\partial r} 
\frac{\partial \rho^2}{\partial r} 
+ 4 \rho^2 W X Y^2 \frac{\partial \psi}{\partial r} 
\frac{\partial W}{\partial r} \right. \\ 
& + & \left. 
2 \rho^4 X Y \frac{\partial \psi}{\partial r} \frac{\partial Y}{\partial r} 
+ 4 \rho^2 V X Y^2 \frac{\partial \psi}{\partial r} 
\frac{\partial Z}{\partial r} 
+ 2 \rho^4 X Y \frac{\partial^2 X}{\partial \theta^2} \right. \\ 
& - & \left. \rho^4 Y \left(\frac{\partial X}{\partial \theta} \right)^2 
+ \rho^2 X Y \frac{\partial X}{\partial \theta} 
\frac{\partial \rho^2}{\partial \theta} 
- \rho^4 X \frac{\partial X}{\partial \theta} 
\frac{\partial Y}{\partial \theta} \right. \\ 
& - & \left. 
\rho^2 Y^2 \frac{\partial X}{\partial r} \frac{\partial \rho^2}{\partial r} 
- \rho^4 Y \frac{\partial X}{\partial r} \frac{\partial Y}{\partial r} 
+ 2 \rho^2 X^2 Y \frac{\partial^2 \rho^2}{\partial \theta^2} \right. \\ 
& - & \left. X^2 Y \left(\frac{\partial \rho^2}{\partial \theta} \right)^2 
- \rho^2 X^2 \frac{\partial \rho^2}{\partial \theta} 
\frac{\partial Y}{\partial \theta} 
+ 2 \rho^2 X Y^2 \frac{\partial^2 \rho^2}{\partial r^2} \right. \\ 
& - & \left. X Y^2 \left(\frac{\partial \rho^2}{\partial r} \right)^2 
+ \rho^2 X Y \frac{\partial \rho^2}{\partial r} \frac{\partial Y}{\partial r} 
- \rho^2 X^2 Y \frac{\partial V}{\partial \theta} 
\frac{\partial Z}{\partial \theta} \right. \\ 
& - & \left. 
\rho^2 X Y^2 \frac{\partial V}{\partial r} \frac{\partial Z}{\partial r} 
- \rho^2 X^2 Y \left(\frac{\partial W}{\partial \theta} \right)^2 
- \rho^2 X Y^2 \left(\frac{\partial W}{\partial r} \right)^2 \right. \\ 
& + & \left. 2 \rho^4 X Y \frac{\partial^2 Y}{\partial r^2} 
- \rho^4 X \left(\frac{\partial Y}{\partial r} \right)^2 \right)
\end{eqnarray*}


\begin{thebibliography}{99}
\bibitem{Berti}
Berti, E., White, F., Maniopoulou, A. \& Bruni, M. 2005
\newblock Rotating neutron stars: an invariant comparison of approximate
and numerical spacetime models.
\newblock {\em \mnras}, {\bf 358}, 923--938. 

\bibitem{Boshkayev}
Boshkayev, K., Quevedo, H. \& Ruffini, R. 2012
\newblock Gravitational field of compact objects in general relativity.
\newblock {\em Phys. Rev. D}, {\bf 86}, 064043 (13 pages). 

\bibitem{Carmeli}
Carmeli, M. 2001
\newblock {\em Classical Fields}.
\newblock World Scientific Publishing.

\bibitem{Castejon}
Castejon-Amenedo, J. \& Manko, V.~S. 1990
\newblock 
Superposition of the Kerr metric with the generalized Erez-Rosen solution.
\newblock {\em Phys. Rev. D}, {\bf 41}, 2018--2020.

\bibitem{Chandrasekhar}
Chandrasekhar, S. 2000
\newblock {\em The Mathematical Theory of Black Holes}.
\newblock Oxford.

\bibitem{Dexter}
Dexter, J. \& Algol, E. 2009
\newblock A Fast new Public Code for Computing Photon Orbits in a Kerr Spacetime.
\newblock {\em \apj}, {\bf 696}, 1616--1629.

\bibitem{Ernst}
Ernst, F.~J. 1968
\newblock New formulation of the axially symmetric gravitational field problem.
\newblock {\em Phys. Rev.}, {\bf 167}, 1175--1177.

\bibitem{Fendt}
Fendt, C. \& Memola, E. 2008
\newblock Formation of relativistic MHD jets: stationary state solutions and numerical simulations.
\newblock {\em International Journal of Modern Physics}, {D 17}, 1677--1686.

\bibitem{Fragile}
Fragile, P.~C., Blaes, O.~M., Anninos, P. \& Salmonson, J.~D. 2007
\newblock Global General Relativistic Magnetohydrodynamic Simulation of a Tilted Black Hole Accretion Disk.
\newblock {\em \apj}, {\bf 668}, 417--429.

\bibitem{Frutos2001}
Frutos-Alfaro, F. 2001
\newblock A computer program to visualize gravitational lenses.
\newblock {\em Am. J. Phys.}, {\bf 69}, 218--222.

\bibitem{Frutos2012}
Frutos-Alfaro, F., Grave, F., M\"uller, T. \& Adis, D. 2012
\newblock Wavefronts and Light Cones for Kerr Spacetimes.
\newblock {\em Journal of Modern Physics}, {\bf 3}, 1882--1890.

\bibitem{Frutos}
Frutos-Alfaro, F., Retana-Montenegro, E., Cordero-Garc{\'{\i}}a, I. \& \\
Bonatti-Gonz\'alez, J.
\newblock Metric of a Slow Rotating Body with Quadrupole Moment from the Erez-Rosen Metric.
\newblock {\em International Journal of Astronomy and Astrophysics}, {\bf 3}, 431-437.
\newblock ArXiv: 1209.6126v2

\bibitem{Hartle}
Hartle, J.~B. \& K.~S.~Thorne, K.~S. 1968
\newblock Slowly Rotating Relativistic Stars. II. 
Models for Neutron Stars and Supermassive Stars.
\newblock {\apj}, {\bf 153}, 807--834.

\bibitem{Hawley}
Hawley, J.F. 2009
\newblock MHD simulations of accretion disks and jets: strengths and limitations.
\newblock {\em Astrophysics and Space Science}, {\bf 320}, 107--114.

\bibitem{Hearn}
Hearn, A.~C. 1999
\newblock {\em REDUCE} (User's and Contributed Packages Manual). 
\newblock Konrad-Zuse-Zentrum f\"ur Informationstechnik, Berlin.

\bibitem{Hoenselaers}
Hoenselaers, C., Kinnersley, W. \& Xanthopoulos, B.~C. 1979
\newblock Symmetries of the stationary Einstein-Maxwell equations. VI. 
Transformations which generate asymptotically flat spacetimes with arbitrary 
multipole moments.
\newblock {\em \jmp}, {\bf 20}(12), 2530--2536.

\bibitem{Kerr}
Kerr, R.~P. 1963
\newblock Gravitational field of a spinning mass as an example of algebraically special metrics.
\newblock {\em Phys. Rev. Lett.}, {\bf 11}, 237--238.

\bibitem{Lewis}
Lewis. T. 1932
\newblock Some Special Solutions of the Equations of Axially Symmetric 
Gravitational Fields.
\newblock {\em Proc. Roy. Soc. Lond.}, A, 176--192.

\bibitem{Manko}
Manko, V.~S. \& Novikov, I.~D. 1992
\newblock Generalizations of the Kerr and Kerr-Newman metrics possessing an 
arbitrary set of mass-multipole moments.
\newblock {\em Class. Quantum Grav.}, {\bf 9}, 2477--2487. 

\bibitem{Manko2000}
Manko, V.~S., Mielke, E.~W. \& Sanabria-G\'omez, J.~D. 2000
\newblock Exact solution for the exterior field of a rotating neutron star.
\newblock {\em Phys. Rev. D}, {\bf 61}, 081501 (5 pages)

\bibitem{Pachon}
Pach\'on, L.~A., Rueda, J.~A. \& Sanabria-G\'omez, J.~D. 2006
\newblock 
Realistic exact solution for the exterior field of a rotating neutron star. 
\newblock {\em Phys. Rev. D}, {\bf 73}, 104038 (12 pages).

\bibitem{Quevedo1986}
Quevedo, H. 1986
\newblock Class of stationary axisymmetric solutions of Einstein's equations 
in empty space.
\newblock {\em Phys. Rev. D}, {\bf 33}, 324--327.

\bibitem{Quevedo1989}
Quevedo, H. 1989
\newblock General static axisymmetric solution of Einstein's vacuum field 
equations in prolate spheroidal coordinates.
\newblock {\em Phys. Rev. D}, {\bf 39}, 2904--2911.

\bibitem{Quevedo1991}
Quevedo, H. \& Mashhoon, B. 1991
\newblock Generalization of Kerr spacetime.
\newblock {\em Phys. Rev. D}, {\bf 43}, 3902--3906.

\bibitem{Quevedo2011}
Quevedo, H. 2011
\newblock Exterior and interior metrics with quadrupole moment.
\newblock {\em Gen. Rel. Grav.}, {\bf 43}, 1141--1152.

\bibitem{Soffel}
Soffel, M.~H. 1989
\newblock {\em Relativity in Astrometry, Celestial Mechanics and Geodesy 
(Astronomy and Astrophysics Library)}.
\newblock Springer-Verlag.

\bibitem{Vincent}
Vincent, F.~H., Paumard, T., Gourgoulhon, E. \& Perrin, G. 2011
\newblock GYOTO: a new general relativistic ray-tracing code.
\newblock {\em ArXiv}: 1109.4769.

\bibitem{Winicour}
Winicour, J., Janis, A.~I. \& Newman, E.~T. 1968
\newblock Static, axially symmetric point horizons.
\newblock {\em Phys. Rev.}, {\bf 176}, 1507--1513.

\bibitem{Young}
Young, J.~H. \& Coulter, C.~A. 1969
\newblock Exact metric for a nonrotating mass with a quadrupole moment.
\newblock {\em Phys. Rev.}, {\bf 184}, 1313--1315. 

\bibitem{Zeldovich}
Zel'dovich, Ya.~B. \& Novikov, I.~D. 2011
\newblock {\em Stars and Relativity}.
\newblock Dover Publications. 
\end{thebibliography}
\end{document}